\documentclass[a4paper,fleqn,usenatbib,useAMS]{mnras}

\usepackage{graphicx}	
\usepackage{amsmath}	
\usepackage{amssymb}	
\usepackage{multicol}        
\usepackage{multirow}
\usepackage{bm}		
\usepackage{pdflscape}	
\usepackage[T1]{fontenc} 
\usepackage{aecompl}

\usepackage[T1]{fontenc}
\usepackage{ae,aecompl}

\title[Building Blocks of the Milky Way]{Building Blocks of the Milky Way's Accreted Spheroid}

\author[P. van Oirschot et al.]{Pim van Oirschot$^{1}$\thanks{e-mail: P.vanOirschot@astro.ru.nl},
Else Starkenburg$^2$, Amina Helmi$^3$ \& Gijs Nelemans $^{1,4}$
\\ 
$^1$Department of Astrophysics/IMAPP, Radboud University Nijmegen, P.O. Box 9010, 6500 GL Nijmegen, The Netherlands \\
$^2$Leibniz-Institut fur Astrophysik Potsdam, An der Sternwarte 16, D-14482 Potsdam, Germany \\
$^3$Kapteyn Astronomical Institute, University of Groningen, P.O. Box 800, 9700 AV, Groningen, The Netherlands \\
$^4$Institute for Astronomy, KU Leuven, Celestijnenlaan 200D, 3001 Leuven, Belgium}

\pubyear{2016}

\begin{document}
\label{firstpage}
\pagerange{\pageref{firstpage}--\pageref{lastpage}}
\maketitle

\begin{abstract}
In the $\Lambda$CDM model of structure formation, a stellar spheroid grows
by the assembly of smaller galaxies, the so-called building blocks.
Combining the Munich-Groningen semi-analytical model of galaxy formation with the
high resolution Aquarius simulations of dark matter haloes, we study the assembly history of 
the stellar spheroids of six Milky Way-mass galaxies, focussing on building block properties 
such as mass, age and metallicity. These properties are compared to those
of the surviving satellites in the same models.
We find that the building blocks have higher star formation rates on average, and this is especially 
the case for the more massive objects. At high redshift these dominate in star formation over 
the satellites, whose star formation timescales are longer on average.
These differences ought to result in a larger $\alpha$-element enhancement from Type II supernovae in the building blocks
(compared to the satellites) by the time Type Ia supernovae would start to enrich them in iron, explaining the observational
trends. Interestingly, there are some variations in the star formation timescales of the building blocks amongst the
simulated haloes, indicating that [$\alpha$/Fe] abundances in spheroids of other galaxies 
might differ from those in our own Milky Way.
\end{abstract}

%
%

\section{Introduction}

The formation and evolution of the Galactic spheroid, consisting of the central bulge and the stellar halo,
has been studied for more than fifty years since the classical paper of \citet{Eggen:1962} on the origin of 
the Milky Way. 
Although it is still unclear to which extent accretion plays a role 
besides instabilities of the disc in the formation of the Galactic bulge \citep[e.g.,][]{Combes:2000,Gerhard:2015,Di-Matteo:2016}, 
there is growing consensus on the formation of the Galactic halo.
Since the proposed scenario of \citet{Searle:1978} in which the stellar halo formed
via the merging of several protogalactic clouds, there have been many pieces of evidence suggesting 
indeed a hierarchical build-up of the Milky Way's stellar halo
\citep[e.g.,][]{Ibata:1994,Helmi:1999a,Belokurov:2006,Bell:2008,Starkenburg:2009,Janesh:2015}. 
Presently, we have a firm theoretical framework provided by the $\Lambda$CDM paradigm predicting 
a hierarchical formation scenario that can be simulated in much detail 
\citep[e.g.,][]{Johnston:1998,Bullock:2001,Bullock:2005,Moore:2006,Abadi:2006}. 
On the other hand, the accretion history of our Galaxy in particular is not completely unravelled yet, 
although much progress is expected thanks to the Gaia mission \citep{Perryman:2001}.
One particularly intriguing question is how the building blocks that formed our Milky Way's accreted spheroid
compare to the satellite galaxies that we see around us today.

In a pioneering paper, \citet{Unavane:1996} attempted to constrain the accretion history of the stellar halo 
from comparisons of the age distribution and chemical abundances of halo stars with those 
of the stars in present-day dwarf spheroidal (dSph) galaxies. Numerous observational studies 
\citep{Shetrone:1998,Shetrone:2001,Shetrone:2003,Tolstoy:2003,Venn:2004,Koch:2008,Tolstoy:2009,Kirby:2010}
reported discrepancies between chemical abundances of satellite galaxies of the Milky Way and field halo stars. 
These studies show that the present-day satellites are, at least partly, unlike the building blocks 
of the Milky Way's stellar spheroid. The dSphs that we see around the Milky Way 
in our Local Group are survivors and thus had naturally more time to form stars than the building block galaxies 
that already dissolved into the halo \citep[e.g.,][]{Mateo:1996}. Even when comparing
equal age populations in both environments \citep[as done by][using RR Lyrae stars]{Fiorentino:2015} 
discrepancies are found between the typical dSphs that survived and those that contributed majorly 
to the build-up of the spheroid. 

In this work we specifically focus on the properties (in terms of mass, age and metallicity) 
of the building blocks of our Milky Way's accreted spheroid modelled within a fully cosmological framework. 
We investigate when they merged and how they relate to the surviving satellite population. 
In the past decade several efforts have already focussed on the build-up of Milky Way stellar haloes 
and/or their chemical evolution, either using hydrodynamical simulations or with semi-analytic techniques 
\citep[e.g.,][]{Bullock:2005,Salvadori:2007,Tumlinson:2006,Tumlinson:2010,Zolotov:2009, 
Cooper:2010, Font:2011,Tissera:2013, Tissera:2014, Cooper:2015, Pillepich:2014, Lowing:2015,Pillepich:2015}. 
A specific focus on chemical evolution has been provided by \citet{Robertson:2005,Font:2006,Font:2006a}, 
using the hybrid semi-analytic plus N-body approach of \citet{Bullock:2005}.
\citet[][C10 hereafter]{Cooper:2010} used the \textsc{galform} semi-analytic galaxy formation model 
to study the disruption of satellite galaxies within the cosmological N-body simulations of the six 
galactic haloes of the Aquarius project \citep{Springel:2008}, which have masses comparable 
to values typically inferred for the Milky Way halo.

We use here a different semi-analytic model to study the formation of our Galaxy and its spheroids'
building blocks than C10 \citep[][hereafter S13, and references therein]{Starkenburg:2013}, 
but using also the Aquarius simulations as a backbone.
In Section~\ref{sec:2} we briefly describe our model, followed by a detailed description 
of the resulting stellar spheroids in Section~\ref{sec:3}.
We will focus on their accreted components, but in this section we will also show how they relate to
the full spheroids in terms of stellar mass.
In Section~\ref{sec:4} we investigate the 
stellar mass $-$ metallicity relation for the building blocks of the accreted spheroids and compare this to the observed 
stellar mass $-$ metallicity relation for the surviving satellite galaxies of the Milky Way, 
and the simulated one by S13.
In this section, we also show that the early star formation (i.e. over 12 Gyrs ago)
in the accreted spheroid was dominated by its building blocks and was much lower in the 
satellite galaxies that survive until the present day.
We apply our analysis to infer observable [$\alpha$/Fe] trends in galaxies 
with various accretion histories in Section~\ref{sec:5} and we conclude in Section~\ref{sec:6}.

Throughout this paper we name all accreted stellar material together the ``accreted spheroid'' of a galaxy. 
This definition is preferred over the term ``halo'' to clarify that this component is present at all radii.
Only in Section~\ref{sec:3}, we furthermore use the term ``accreted bulge'' for the innermost 3~kpc of the
accreted spheroid.

\section{The Model}\label{sec:2}
We use the semi-analytic model for galaxy formation that was originally established
in Munich \citep{Kauffmann:1999,Springel:2001,De-Lucia:2004,Croton:2006,De-Lucia:2007,De-Lucia:2008} 
and developed further in Groningen \citep[][S13]{Li:2010}.
The merger history trees of the six Milky Way-like haloes of the Aquarius project, denoted A $-$ F \citep{Springel:2008},
and their substructures were constructed using the \textsc{subfind} algorithm \citep{Springel:2001},
after which baryonic processes are modelled using simple but observationally and 
astrophysically motivated prescriptions \citep[][S13, and references therein]{De-Lucia:2004,Li:2010}.

A galaxy merger tree was constructed to follow the galaxies that end up in the Milky Way's stellar spheroid over time.
Each building block of the spheroid undergoes three phases in this galaxy merger tree: a first phase where it is a main galaxy
with its own dark matter halo; a second phase where it is a satellite galaxy (its dark matter halo becomes a subhalo of a more massive halo);
and a so called orphan phase, where the dark halo of a satellite galaxy is tidally stripped down to below the \textsc{subfind} resolution 
limit of 20 particles. Up until this last point, where the galaxy has `lost' its dark matter halo,
the galaxy merger tree is identical to the dark matter merger tree.

As explained in detail by \citet{De-Lucia:2008}, stellar spheroids grow via galaxy mergers and disc instabilities in our model. 
In situ star formation only takes place in discs of galaxies, not in spheroids.
However, during a major merger, the disc of the galaxy gets completely destroyed and all its stars end up in the spheroid, 
including the stars that just have been formed in the starburst initiated by the collision. This could be regarded as in situ 
star formation in the spheroid.
We classify a galaxy merger as \emph{major} if the mass ratio (mass in stars and cold gas) of the merging
galaxies is larger than 0.3. 
In a \emph{minor} merger, for which this mass ratio of the merging galaxies is smaller than or equal to 0.3, 
the stars of the least massive galaxy are added to the spheroid of the more massive one, thereby leaving the disc of the latter intact. 
Whenever a galaxy is disc-dominated (spheroid stellar mass / total stellar mass $< 0.1$) and meets the disc instability condition
\begin{equation}
\frac{V_\mathrm{max}}{(Gm_\mathrm{disc}/r_\mathrm{disc})^{1/2}} \leq 1.1
\end{equation}
\citep{Efstathiou:1982, Mo:1998}, half the disc mass is transferred to the spheroid in our code to make the disc stable again. 
In this equation $V_\mathrm{max}$ is the maximum velocity of the main halo, $m_\mathrm{disc}$ and $r_\mathrm{disc}$ are the
stellar mass and the radius of the disc respectively, and $G$ is the gravitational constant. 

Despite the semi-analytical nature of our model in which no stellar particles are explicitly modelled or tagged, 
our model includes some prescription of stellar stripping in merging satellites, but only when the dark matter halo
of a satellite galaxy is so heavily stripped that its half-mass radius becomes smaller than the half-mass radius 
of the stars and cold gas. In this case, the stars and cold gas are removed up to
the half-mass radius of the dark matter and added to the host spheroid (see S13, Appendix A1). 
Orphan galaxies - a class that is particularly difficult to handle well, because no information on them is present 
in the simulations anymore - either merge with the central galaxy on a dynamical friction timescale
before redshift zero, or, if this timescale is longer, might survive. In the latter case, 
their survival depends on their average mass density compared to that of their host system (see S13, Appendix A2).
We note that besides stripping of their dark matter mass and possibly some stellar content, 
surviving orphan galaxies are similar to any surviving dwarf galaxy including a chemical history of self-enrichment 
\citep[which sets them apart from globular clusters for instance,][]{Kruijssen:2012,Leaman:2012,Willman:2012}. 
Some fainter orphan galaxies could potentially even still be dark matter dominated.
Note that the tidal disruption of orphan galaxies is another way of growing the spheroid, as is stellar stripping.

To summarize, the five ways of spheroid growth in our model are: (1) major mergers, (2) minor mergers on a dynamical friction timescale, (3) disc instabilities,
(4) stellar stripping and (5) tidal disruption of orphans. The four different types of galaxies that we model are: (1) Main galaxies that do survive until the
present day. These galaxies have a dark matter halo and may have several satellite galaxies in subhaloes of 
this dark halo or orphan galaxies bound to them. The most massive of these main galaxies in our simulation is our model Milky Way. 
(2) Building block galaxies that once were main galaxies but went through the above-mentioned 
three phases of evolution before merging with our progenitor Milky Way galaxy. These do not survive until the 
present day. (3) Surviving satellite galaxies of a main galaxy, which do have a dark matter halo that is a
subhalo of the dark halo containing the main galaxy.
(4) Orphan galaxies that might eventually merge with a main galaxy but survive until the present day.

Our model assumes an instantaneous recycling approximation, i.e. we do not take into account finite stellar lifetimes.
The abundance of an $\alpha$-element such as Mg for instance, mainly originating in short-lived supernovae (SNe) type II, 
can therefore better be compared with the metallicity predicted by our model than Fe, which is thought to originate 
mainly in the (delayed) population of type Ia SNe.
No individual elements are explicitly traced in the model though, instead it returns a total mass in metals for any system 
in the gas and stars. Throughout this paper, we refer to the metallicity of any stellar system by 
$\log [\mathrm{Z}_\mathrm{stars}/\mathrm{Z}_\odot]$, i.e.
the logarithm of the ratio of mass in metals over the total mass in stars in that galaxy, divided by the solar metallicity, 
$\mathrm{Z}_\odot = 0.02$. Two discrepancies however arise when comparing this modelled metallicity with observational data: 
(1) often in observational data [Fe/H] is measured, rather than [Mg/H], and (2) the average [Fe/H] of a stellar system is 
calculated by taking the average of the logarithms of (a representative sample of) its individual stars' iron abundance ratios 
compared to that of the Sun, whereas we take the logarithm of the average metallicity compared to that of the Sun. 
Thus the metallicity values that we find are higher.
In our comparisons with data we attempt to compensate for both discrepencies. 
We refer the reader to S13 for details (i.e. see their Equation~2 and the paragraph below that equation), 
but in short we convert each measured [Fe/H] into [Mg/H] and apply an offset 
to correct for the difference in the averaging procedure.

\section{The Stellar Spheroids}\label{sec:3}

Our modelled stellar spheroids are part of galaxies that were already analysed in some detail in S13, 
who looked at the total stellar mass of the galaxies that developed 
in the six Aquarius haloes as well as their bulge/disc ratios. 
As pointed out by S13, if viewed as one component, the integrated or average values 
for the spheroids can best be compared with the Milky Way bulge
since the stellar halo contains very few stars compared to the bulge.
In terms of metallicity, spheroid B has the closest match to the observed bulge
metallicity of the Milky Way of [Fe/H]~$\sim-0.25$ \citep{McWilliam:1994,Zoccali:2003,Bensby:2013}.
S13 found that the bulge/disc ratios of the simulated galaxies in all Aquarius haloes
except E and F are close to the estimated value of $0.2-0.3$ \citep{Bissantz:2004} for the Milky Way. 
We list the disc stellar mass and the spheroid stellar mass in Table~\ref{tab:1}.
The simulated galaxies in Aquarius haloes B and E are the closest Milky Way analogs 
in terms of total stellar mass, i.e. the sum of these first two columns.

\begin{figure*}
 \includegraphics[width=\textwidth]{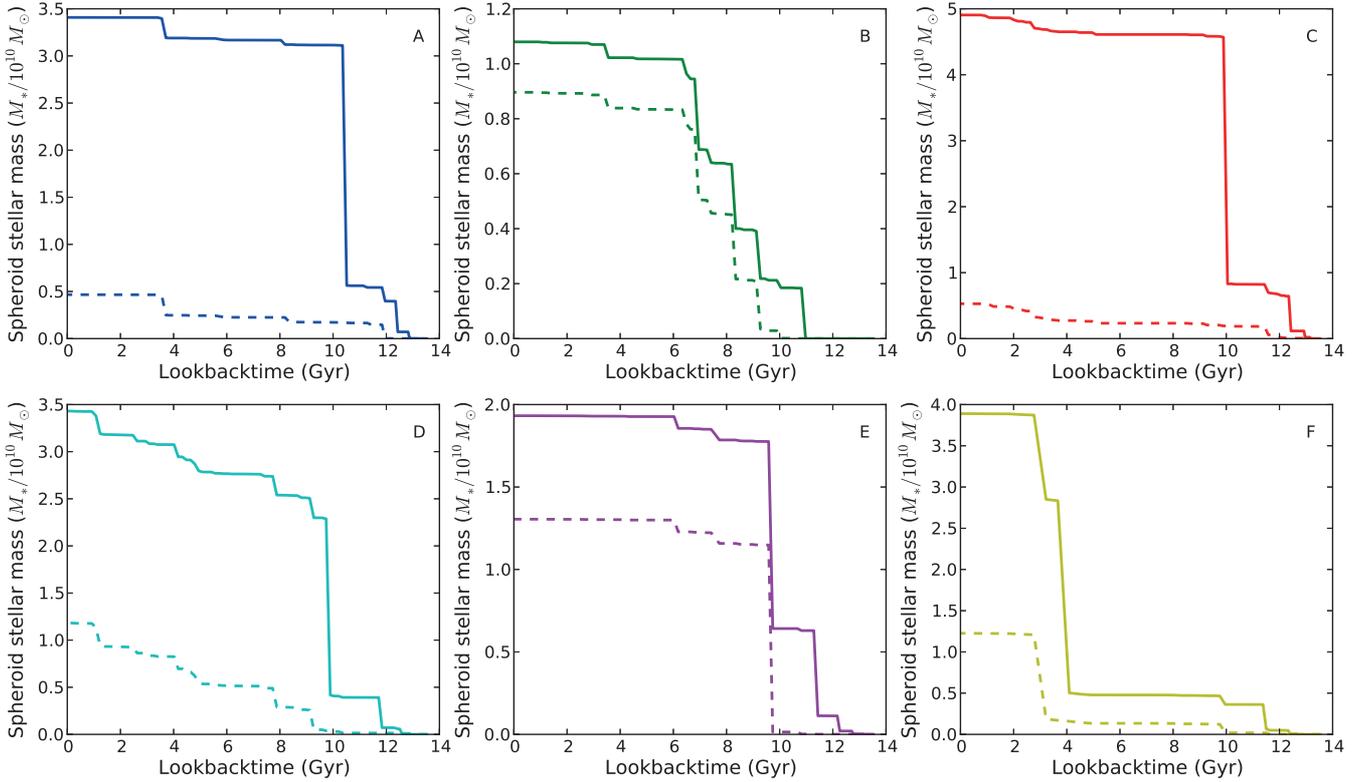}
 \caption{Build-up of the six Aquarius spheroids in stellar mass over time.
 Solid lines indicate the full spheroid growths, including stars that were moved
 from the disc to the spheroid by the disc instability channel (see Section~\ref{sec:2}). 
 Dashed lines indicate the spheroid growth by accretion only.
 }
 \label{fig:1}
\end{figure*}

In Figure~\ref{fig:1} we show the growth of the six Aquarius spheroids over time, including the stars
that were moved to the spheroid through disc instability (solid lines), or excluding this and showing the contribution 
from accreted stellar mass only (dashed lines). 
The final spheroid masses can be read off from the vertical axis of each panel (and are listed in 
Table~\ref{tab:1}). They vary from $1.1 \times 10^{10} M_\odot$ for spheroid B to
$5.2 \times 10^{10} M_\odot$ for spheroid C. Whereas spheroid B grows more gradual, 
the other Aquarius spheroids have undergone one or two major growths. 
Interestingly, whereas Aquarius haloes B and E contain the least massive 
galaxies in terms of total stellar mass (S13), they have the largest fraction of accreted spheroid stars 
compared to their total spheroid mass, since the contribution of disc instabilities
to the build-up of the total spheroid is smallest in these haloes.
The fraction is smallest in Aquarius haloes A and C, which also have the 
lowest accreted stellar mass in an absolute sense.
Whereas the disc instability channel is mainly considered to lead to the formation of the galactic bar 
that is much more centrally concentrated, accreted material will contribute at all radii. 
The largest mass ratio (mass in stars and cold gas) that we find for galaxies merging with the central galaxy
in our simulation is 0.23. Because we assign the label \emph{major} only to a merger in which this ratio is 
larger than 0.3, none of the Milky Way galaxies in our simulations have undergone a major merger during their 
lifetimes, and even their mergers with the most massive building blocks are classified as minor.
Consequently, we do not find any stars that were formed in situ in our spheroids, since
in situ star formation only happens in major mergers in our model.

\begin{table}
 \caption{The stellar mass of the disc, total spheroid stellar mass, the accreted spheroid stellar mass,
 the fraction of accreted spheroid stellar mass contributed by surviving satellite galaxies
 $f_\mathrm{surv}$, the percentage of surviving satellites that is orphan,
 and the number of significant progenitors $N_\mathrm{prog}$, per halo.}
 \begin{tabular}{ccccccc}
  \hline
  \multirow{2}{*}{Halo} & \underline{ $M_{*,\mathrm{disc}}$} & \underline{$M_{*,\mathrm{sph.}}$} & \underline{$M_{*,\mathrm{acc.}}$} & \multirow{2}{*}{$f_\mathrm{surv}$} & orph & \multirow{2}{*}{$N_\mathrm{prog}$}\\
  & ${}_{10^{10} M_\odot}$ & ${}_{10^{10} M_\odot}$ & ${}_{10^{10} M_\odot}$ & & \% &\\
  \hline 
  A & 15.35 & 3.409 & 0.467 & 0.104 & 36 & 3.3\\
  B & 5.925 & 1.080 & 0.896 & 0.034 & 23 & 3.1\\
  C & 15.79 & 4.906 & 0.529 & 0.482 & 45 & 6.1\\
  D & 12.40 & 3.431 & 1.182 & 0.484 & 32 & 5.4\\
  E & 3.156 & 1.932 & 1.305 & 0.009 & 35 & 1.3\\
  F & 5.182 & 3.890 & 1.226 & 0.896 & 31 & 1.4\\
 \hline
 \end{tabular}\label{tab:1}
\end{table}

In Table~\ref{tab:1} the stellar masses of these accreted spheroids are listed, as well as the fraction of the accreted 
spheroid stellar mass that is material stripped from surviving satellites $f_\mathrm{surv}$. 
This fraction is very large for Aquarius halo F, because the progenitor that almost
completely accounts for its spheroids total stellar mass has a still surviving counterpart.
For Aquarius haloes C and D we also find large values of $f_\mathrm{surv}$, which is in agreement 
with the result from C10. 
The difference between our results and those of C10 - in particular for halo F, 
but in lesser extent for the other haloes - is probably due to our differences in 
the treatment for the tidal disruption of orphan galaxies and stellar stripping 
(C10 make use of a particle tagging technique).

Table~\ref{tab:1} also lists the percentage of surviving satellites that is orphan, when
using the S13 prescriptions that we summarized in Section~\ref{sec:2}. We see that they
constitute $\sim 1/3$ of the total population of surviving satellites in most haloes.

The stellar mass of the accreted spheroids is on average $9.3 \cdot 10^{9} \ \mathrm{M}_\odot$ in our models. 
The resulting accreted spheroid/disc mass ratios that we find are 0.03, 0.15, 0.03, 0.10, 0.41 and 0.24 for haloes 
A$-$F respectively. Using the same Aquarius haloes but a different semi-analytic model, C10 
find on average accreted spheroids of $1.2 \cdot 10^9 \ \mathrm{M}_\odot$, thus on average almost a factor 8 lower. 
To find differences between the two codes in these values is not so surprising; the stellar mass $-$ halo mass relation in our model is quite 
different from that in \textsc{galform} in this mass range due to their stronger feedback (see S13 for a discussion).

\begin{figure*}
\includegraphics[width=\textwidth]{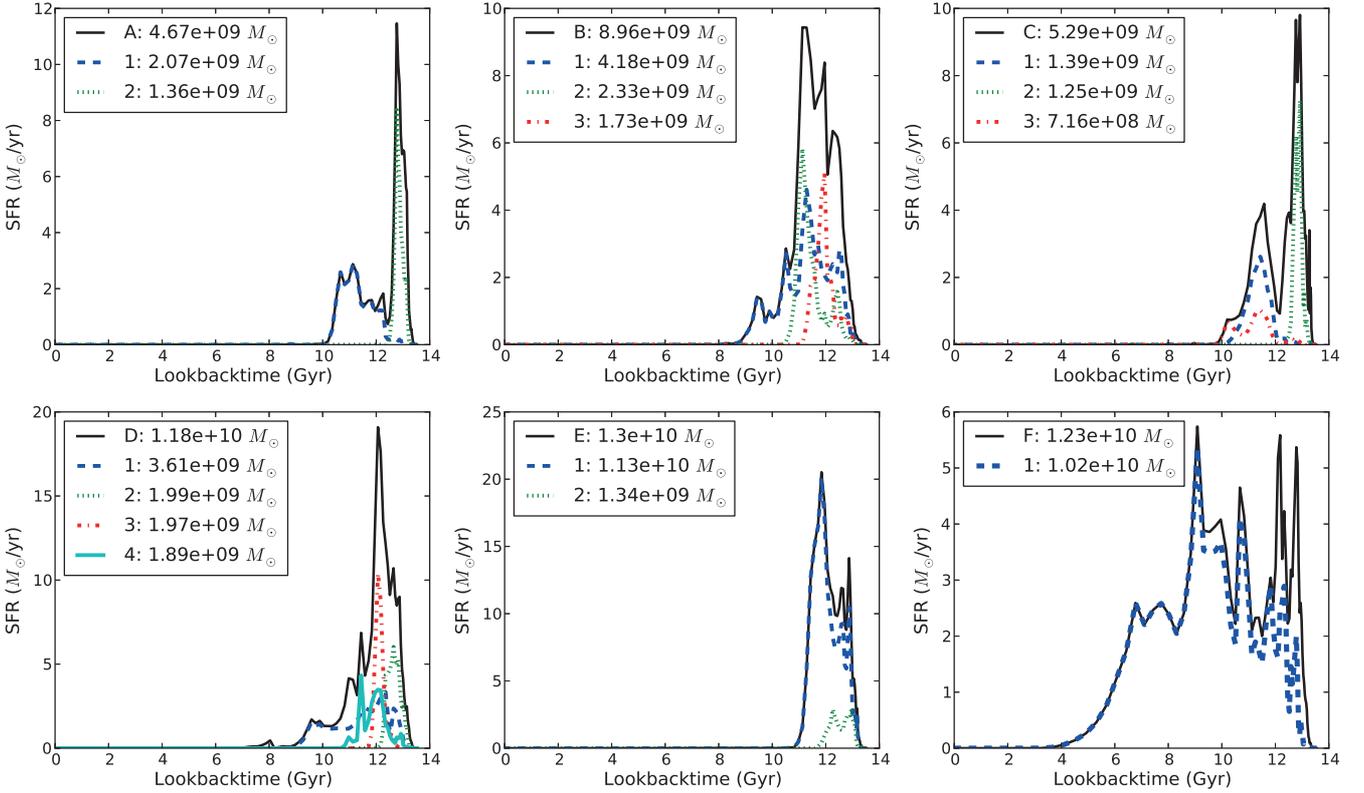}
 \caption{Star formation rates of the accreted spheroids (black solid lines),
 in combination with the star formation rates of their main progenitor galaxies, 
 i.e. the largest building blocks (dashed, dotted and dot-dashed lines) that contribute to them. 
 These main progenitors were selected to contribute at least 10\% of the spheroid's stellar mass.
 The first number in the legend of each panel is again the total accreted stellar mass of that spheroid,
 followed by the stellar mass of the largest building blocks.\newline}
\label{fig:2}
\end{figure*}

An observational estimate of the Galactic halo's stellar mass can be made from the local halo mass density
$\rho_0 = 1.5 \cdot 10^{-4} \ \mathrm{M}_\odot \ \mathrm{pc}^{-3}$ \citep{Fuchs:1998}
combined with the density function 
\begin{equation}
\rho(x,y,z) = \frac{\rho_0}{r_0^n}\left(x^2 + y^2 + \frac{z^2}{q^2}\right)^{n/2} \label{eq.1}
\end{equation}
\citep{Helmi:2008}, where $r_0 = 8.0$~kpc \citep{Reid:1993}
is the distance from the Sun to the Galactic centre, $q=0.64$ the minor-to-major axis ratio and 
$n=-2.8$ the power law exponent of the density profile \citep{Juric:2008}.
Assuming this mass density profile holds for the stellar halo from a 3~kpc distance of the Galactic centre
out to 20~kpc, an integration between these boundaries yields $1.2 \cdot 10^9 \ \mathrm{M}_\odot$
as an analytical estimate of the stellar halo mass.
Beyond 20~kpc the stellar mass density slope $n$ steepens \citep{Bell:2008,Deason:2014}
and the extra stellar mass that is obtained by integrating further is similar to the uncertainty
caused by errors on the estimated values of the parameters used for the integration in the range $3-20$~kpc.

On average 38\% of the accreted spheroid stars in the \textsc{galform}
semi-analytic model of C10 is located in the innermost 3~kpc. They refer to this component as the accreted bulge. 
The spread between these fractions in the six Aquarius haloes is large within their model, 
i.e. 26, 59, 8.7, 12, 90 and 20\% for haloes A$-$F respectively. 
Since we have no particle tagging scheme implemented, we can not make this distinction
based on distances in our models, however if the accreted spheroid stars 
in our model were distributed among bulge (inner 3~kpc) and halo according to the percentages derived from C10, 
the accreted stellar haloes (excluding the accreted bulge) A$-$F would respectively be 
3.4, 3.6, 4.8, 10, 1.2 and 9.7 $\cdot 10^9 \ \mathrm{M}_\odot$ in our model. 
Aquarius halo E, containing the lowest mass galaxy, also contains the lowest mass stellar halo.
Note that this estimate of its stellar halo mass matches our analytical estimate above. 

The rightmost column in Table~\ref{tab:1} lists the number of significant progenitors $N_\mathrm{prog}$
of the accreted spheroids. Following C10, this number is defined as 
$N_\mathrm{prog} = M_{\mathrm{tot},*}^2/\sum_i m_{\mathrm{prog},i,*}^2$, which is the total number of progenitors 
in the case where each contributes equal mass, or the number of significant progenitors in the case where the remainder 
provide a negligible contribution. $m_{\mathrm{prog},i,*}$ is the total stellar mass of a building block, 
or the total stellar mass stripped from one and the same surviving satellite.
The sum of all progenitor masses thus equals the total accreted stellar mass
of the spheroid ($\sum_i m_{\mathrm{prog},i,*} = M_{\mathrm{tot},*}$).
While the full spheroids (including the disc instability mechanisms) are dominated by one or two major growths only, 
as shown in Figure~\ref{fig:1}, we see from Table~\ref{tab:1} that the accreted spheroids are typically built out of several 
significant building blocks.
 
We find that the stellar spheroid is built almost completely by a few main progenitor galaxies,
as C10 find for the stellar halo. However, the number that we find is for all Aquarius haloes, 
except halo A, higher than the result of C10. 
Most significantly for halo C (for which they find $N_\mathrm{prog} = 2.8$), although only marginally for haloes E and F.
For halo C, we find 9 building blocks with a stellar mass larger than $10^8 M_\odot$, 
which is more than in any of the other Aquarius haloes, and indeed points to a large number of significant progenitors,
even though the total accreted mass of this spheroid is one of the lowest (see Table~\ref{tab:1}). 
Our accreted spheroids B, E and F have the lowest number of progenitors, in agreement with 
C10's finding that the accreted haloes corresponding to these Aquarius haloes have the lowest $N_\mathrm{prog}$.
The larger number of significant progenitors that we find for haloes B, C and D could 
be due to our different selection of radii, i.e. C10 do not include any stars within a 3~kpc distance from the 
Galactic centre in their halo selection.
The other differences  between our model and that of C10 that could again play a role
are the different stellar mass $-$ halo mass relation, and the treatment of 
stellar stripping and the tidal disruption of orphan galaxies.

\begin{figure*}
 \includegraphics[width=\textwidth]{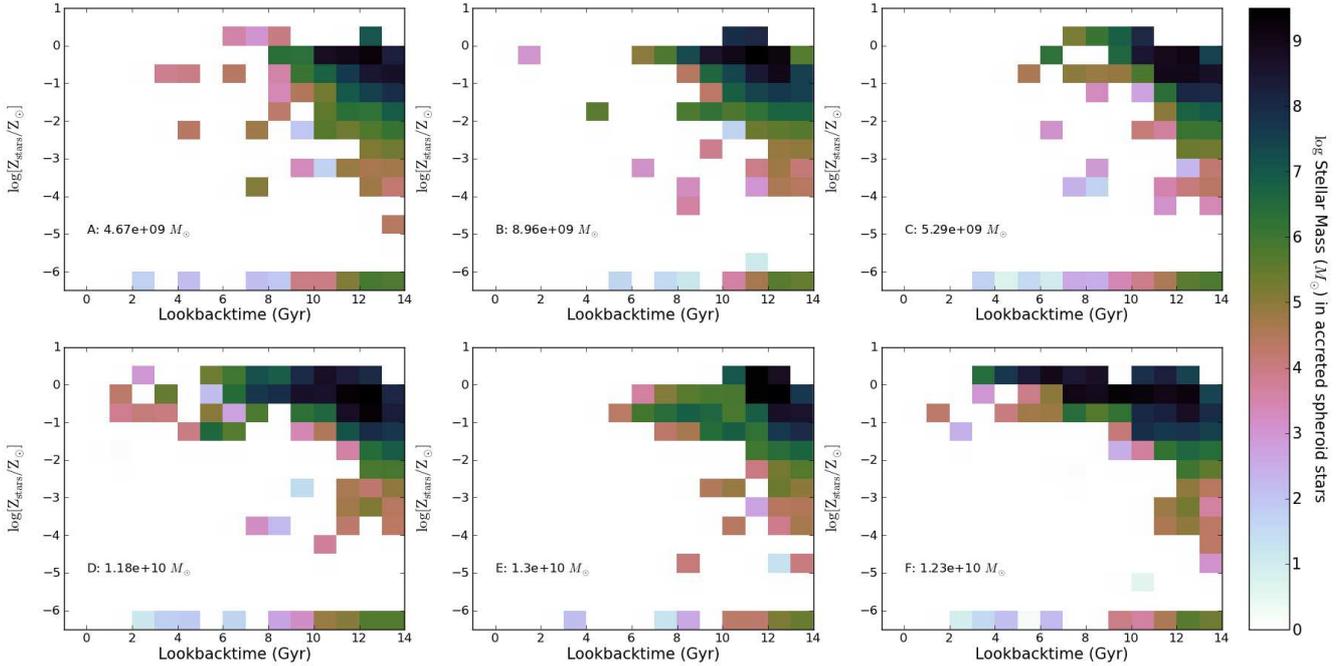}
 \caption{Age-Metallicity Maps ($\log [\mathrm{Z}_\mathrm{stars}/\mathrm{Z}_\odot]$) 
 of the six accreted spheroids.
 The colormap represents the stellar mass ($M_\odot$) per bin, on a logarithmic scale.
 In the bottom left corner of each panel the total accreted stellar mass of that spheroid is indicated.}
 \label{fig:3}
\end{figure*}

Figure~\ref{fig:2} shows the spheroids' star formation rates (SFRs, in $M_\odot$/yr) as a function of 
lookbacktime. The six black solid lines in Figure~\ref{fig:2} show the total SFRs of the
accreted spheroids, i.e. the sum of the SFRs of all building blocks. With dashed, dotted, dot-dashed 
and coloured solid lines the SFRs of the main progenitor galaxies are shown.
These main progenitors were selected to contribute at least 10\% of the spheroid's stellar mass.
Note that this selection criterion is more simplistic than the non-integer statistical count 
of main progenitors as defined in C10 and presented in Table~\ref{tab:1}.
The blue dashed lines represent the SFRs of the most massive progenitor galaxies, the green dotted
lines the second most massive progenitors, red dot-dashed the third, and cyan solid the fourth
(if applicable). 
Most of these main progenitors are completely disrupted, but for haloes C, D and F, respectively 69\%, 94\% 
and 54\% of the original stellar mass of a still surviving satellite constitutes its most massive ``building block''.

In Figure~\ref{fig:3} we show the ages (in bins of 1 Gyr) and metallicities ($\log [\mathrm{Z}_\mathrm{stars}/\mathrm{Z}_\odot]$, 
in bins of 0.5 dex) of the accreted spheroids. 
The stars in the lowest luminosity bin ($\log [\mathrm{Z}_\mathrm{stars}/\mathrm{Z}_\odot]< -6$) in Figure~\ref{fig:3} 
were formed without any metals; that these (still) exist in our model is due to the neglection of any kind of pre-enrichment 
from Population III stars in our model.
The majority of our modelled accreted stellar spheroid population have near solar metallicity abundances.
This is in agreement with the observed average [Mg/H] value for bulge stars,
eg. [Fe/H]~$\sim-0.25$ \citep{McWilliam:1994,Zoccali:2003,Bensby:2013}, which combined with [Mg/Fe]~$\sim+0.25$
brings the observed average [Mg/H] to solar.
However, it is difficult to compare our model, which predicts the metallicity distribution function of spheroid stars 
over all radii, directly to an observed one \citep[eg.][]{An:2013,Allende-Prieto:2014}, because the observations are made in a certain direction and/or distance range. 
If we interpret the lower metallicity accreted spheroid stars as halo stars, we seem to underestimate the lowest metallicities ($<\sim-2$) 
compared to the observed distributions by \citet{An:2013} and \citet{Allende-Prieto:2014}.

From Figures \ref{fig:2} and \ref{fig:3}, it is clear that the majority of spheroid stars is old. 
An exception is the stellar spheroid of F, but this galaxy is atypical as a Milky Way analog because its
spheroid was built mainly by the recent stripping of stars from one still surviving (orphan) satellite,
i.e. 3 Gyr ago (see the bottom right panel of Figure~\ref{fig:1}).

As expected, our total spheroid's SFR values are very similar to those presented in Figure~8 in \citet{De-Lucia:2014}
who use a slightly different version of the same semi-analytic code. Spheroid~F looks very different between both models, 
but this is mainly due to the single massive progenitor galaxy mentioned above that either is counted as part of the 
Galactic spheroid (in our model) or not (in their model). 
In the corresponding erratum, \citet{De-Lucia:2015} moreover show that when they apply a new technique to account for finite stellar lifetimes in 
their model, rather than relying on an instantaneous recycling approximation, the total spheroid masses are typically
lower by a factor of $\sim$2. The implementation of finite stellar lifetimes does also have a significant effect on their 
overall spheroid metallicity; these are typically lowered by $\sim$0.5 dex.

\section{Comparison of Building Blocks and Surviving Satellites: Metallicity and SFR}\label{sec:4}

Having presented the SFRs of the most massive building blocks in the previous section and Figure~\ref{fig:2},
we will now discuss in more detail the properties of all the building block galaxies 
and compare them with those of the surviving satellites. We would like to point out that although the number of building blocks
is smaller than the number of surviving satellites (for spheroid~E the number of building blocks is even less than half the number
of surviving satellites), the total stellar mass in building blocks is much larger than the stellar mass in
surviving satellites for most spheroids, for spheroid B even more than a factor 10. The exceptions are spheroid A 
(where the stellar mass is similar in the two populations) and spheroid F (which has more mass in surviving satellites).

In their Figure~5, S13 showed the luminosity-metallicity relation for the 
satellite galaxies of all Aquarius haloes and concluded from a comparison with observed average [Mg/H] values
for the Milky Way satellites that the model resembles reality quite accurately. 
An exception are the model galaxies with metallicities~$< -3$. Many more of those are seen in the models than we observe 
around the Milky Way. However, as explained in S13, these typically have experienced star formation in less than 4 
snapshots of our simulation, resulting in a clear signature of the neglect of pre-enrichment from the very first 
generation of stars. All stars formed in the first star formation event are 
modelled to have no metals at all and an identical initial mass function (IMF) to more metal-rich components. 
A different, more top-heavy, IMF for these first stars is however likely to enrich these galaxies easily to a 
metallicity floor of [Fe/H]~$\sim - 3$ \citep[e.g.,][]{Salvadori:2008}. 

\begin{figure}
 \includegraphics[width=\columnwidth]{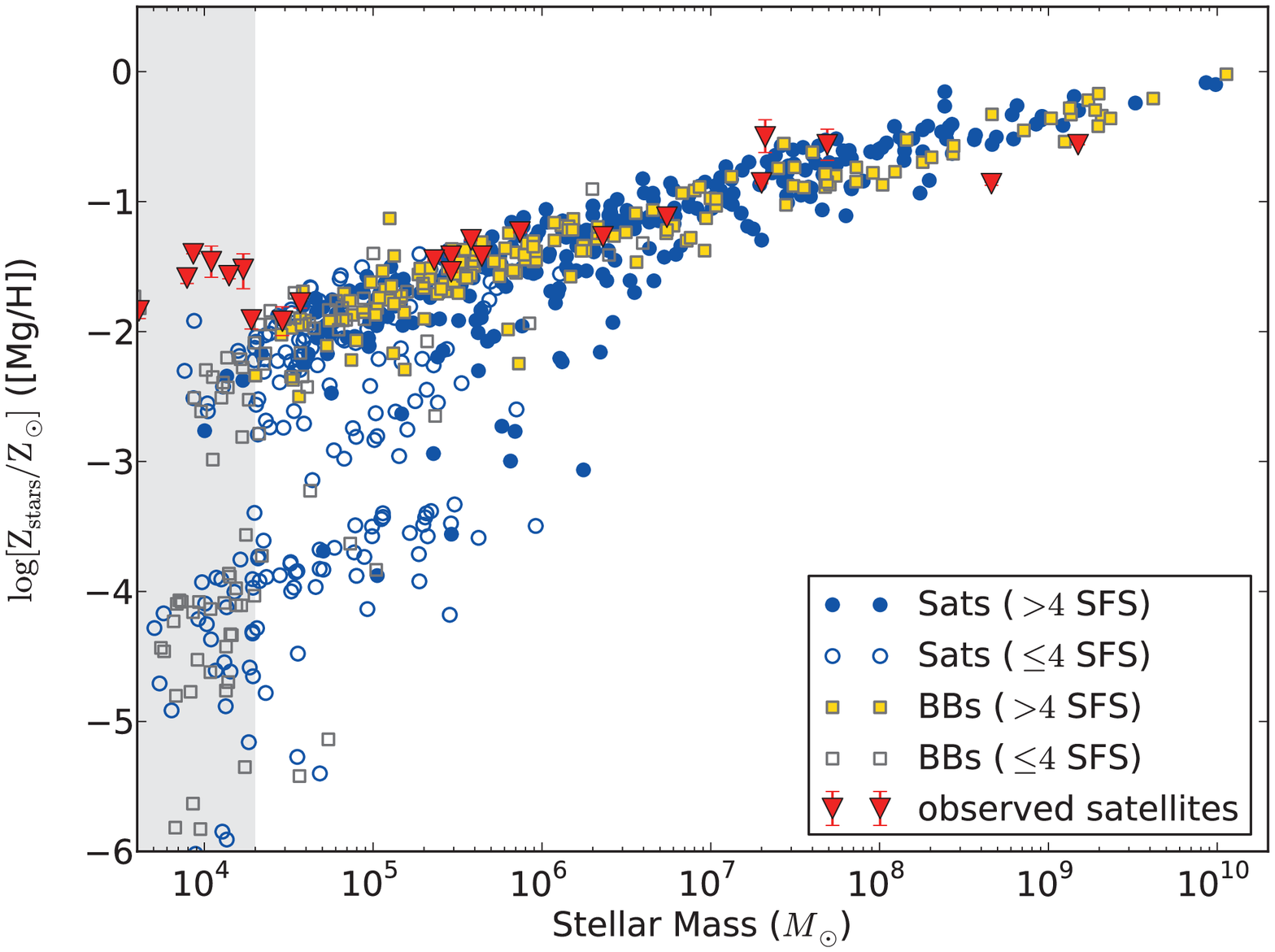}
 \caption{Stellar mass versus metallicity ($\log [\mathrm{Z}_\mathrm{stars}/\mathrm{Z}_\odot]$) relation for the building blocks
 (yellow boxes) and for the surviving satellites (blue circles). Building blocks/surviving satellites 
 with more than 4 snapshots during which there is star formation (star formation snapshots, or SFS)
 are shown with filled marker symbols, those with less than or equal to 4 SFS with open symbols.
 The 88 ``building blocks'' that are stripped material of a surviving satellite (17\% of the total number of BBs) are not shown.
 With a grey zone we indicate what mass range of the building blocks/surviving satellites we 
 do not trust due to the limiting resolution of our simulation.
 Observed satellite metallicity values \citep{McConnachie:2012}, corrected to approximate a mass-weighted average of [Mg/H] 
 for a better comparison to the models (see text for details) are plotted as red triangles with errorbars.
 The seven points in the gray area correspond to Leo IV, Bootes III
 (which nature is unclear), Ursa Major (I), Leo V, Pisces II, Canes Venatici II and Ursa Major II, 
 from right to left respectively. }
 \label{fig:4}
\vspace{1cm}
 \includegraphics[width=\columnwidth]{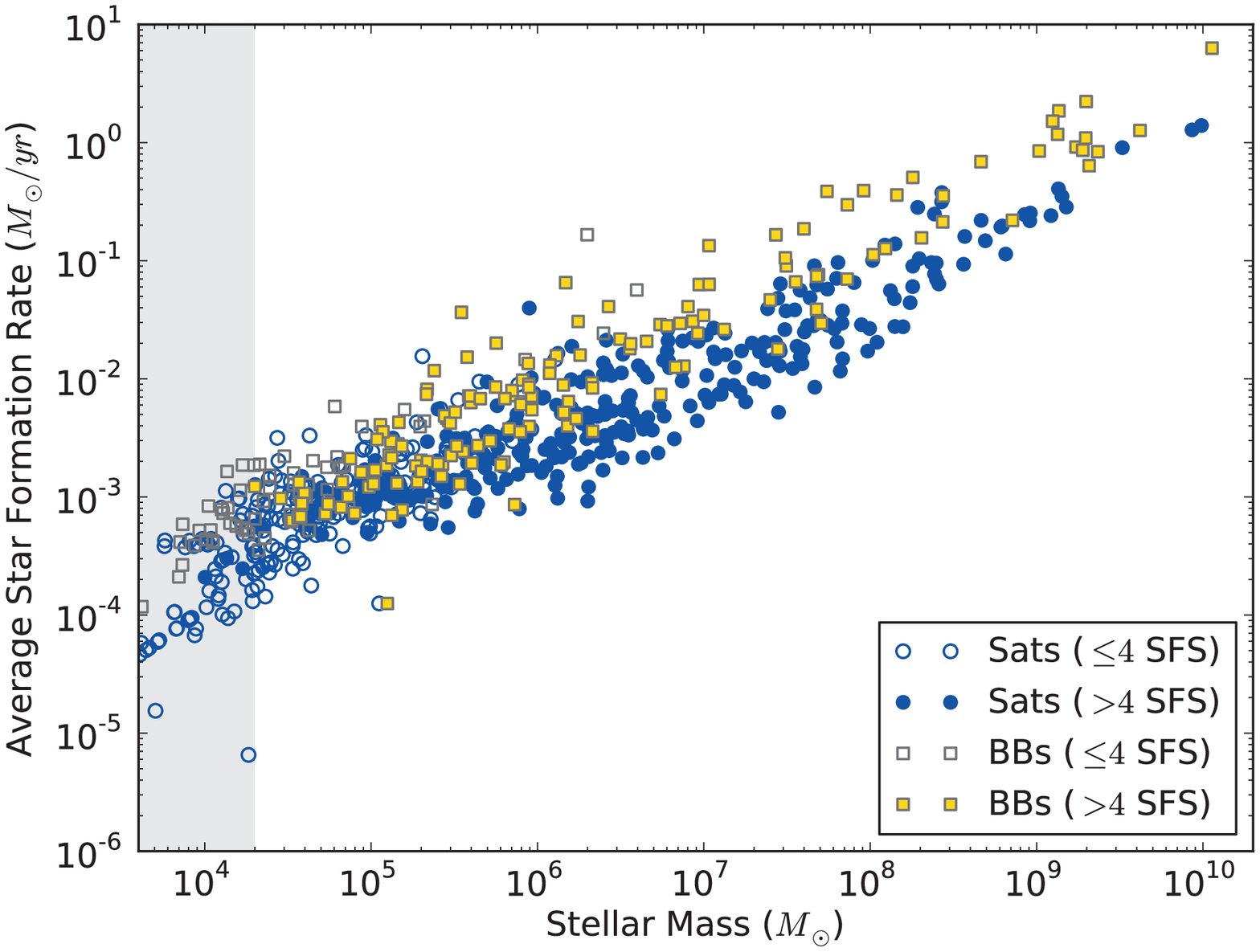}
 \caption{Stellar mass versus the average Star Formation Rate for the building blocks
 (yellow boxes) and for the surviving satellites (blue circles). Building blocks/surviving satellites 
 with more than 4 SFS are shown with filled marker symbols, those with less than or equal to 4 SFS with open symbols.
 The 88 ``building blocks'' that are stripped material of a surviving satellite (17\% of the total number of BBs) are not shown.
 With a grey zone we indicate what mass range of the building blocks/surviving satellites we 
 do not trust due to the limiting resolution of our simulation.}
 \label{fig:5}
 \end{figure}

\begin{figure*}
\includegraphics[width=\columnwidth]{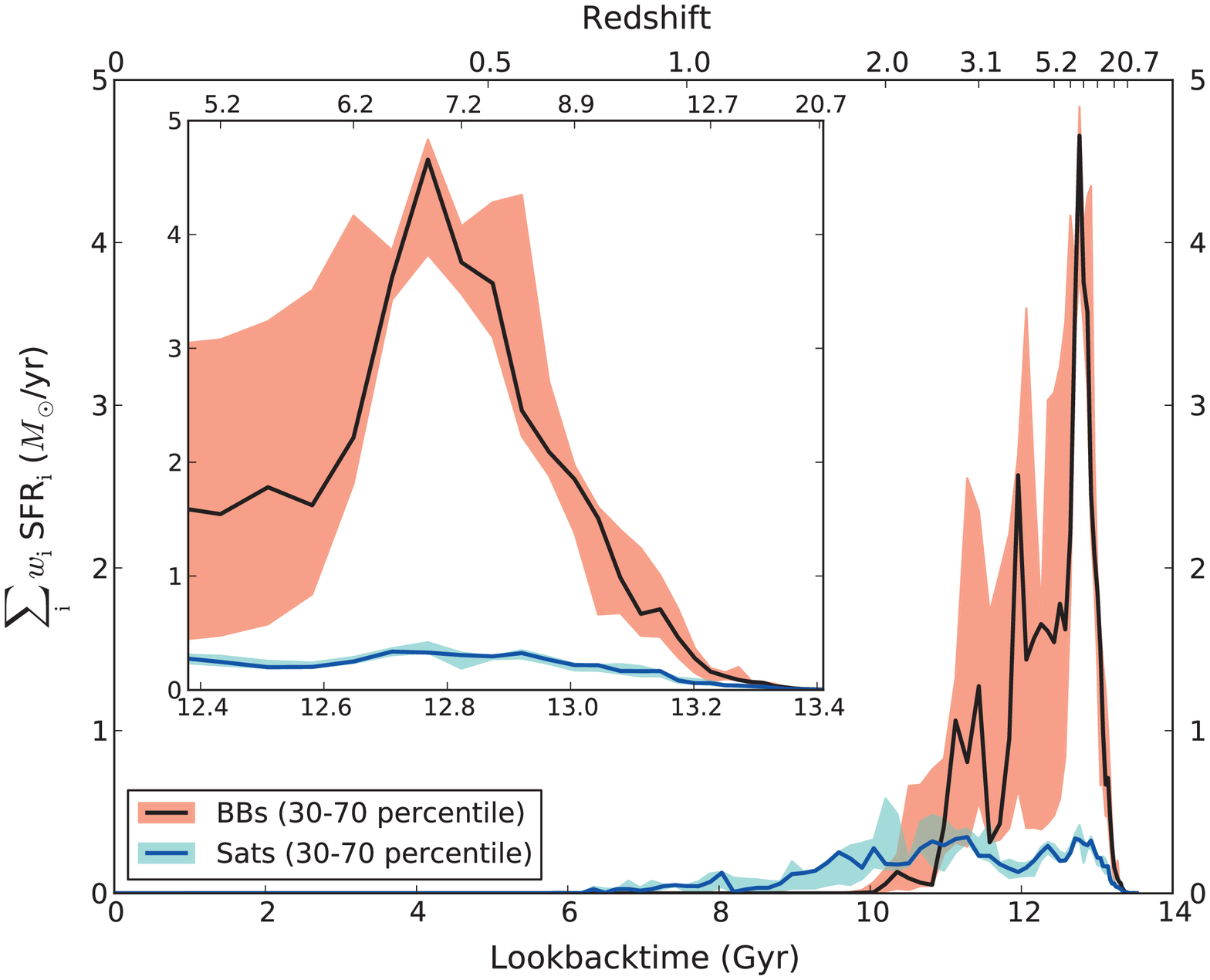}
\includegraphics[width=\columnwidth]{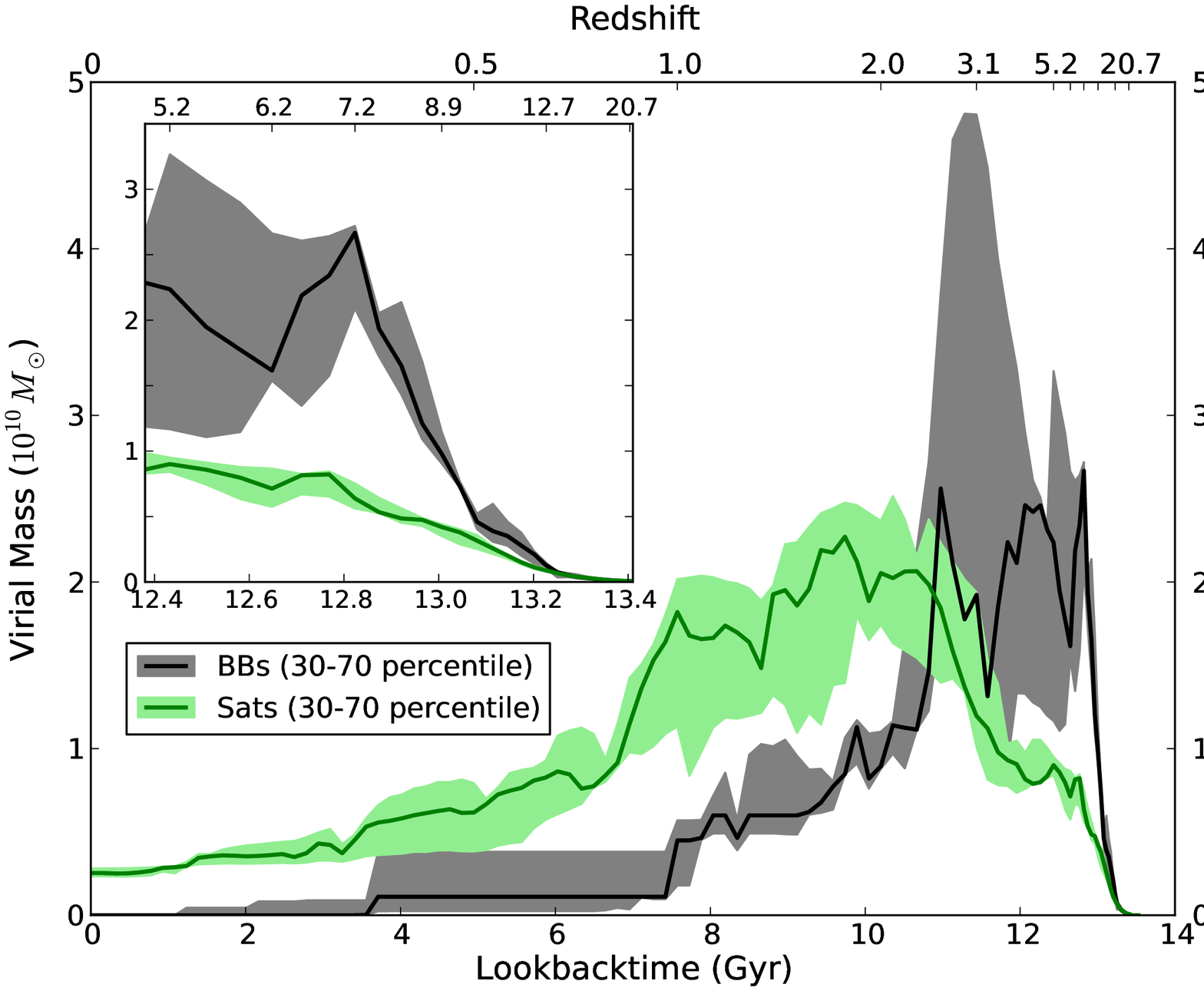}
 \caption{Left panel: The star formations rates ($M_\odot$/yr) of the fully disrupted building blocks (black line with red band)
 and the surviving satellites (blue line with light blue band) in haloes A$-$E, weighted by the mass 
 of that galaxy at that time of star formation. At each time, the sum of the weights of the fully disrupted building blocks
 as well as those of the surviving satellites add up to one for each of these five Aquarius haloes ($\sum_\mathrm{i} w_\mathrm{i} = 1$),
 after which the median is plotted (black and blue lines) as well as the 30$-$70 percentiles (red and light blue bands).
 The corresponding redshift is shown at the top axis. In the zoom-in panel the SFRs in the first Gyr are shown.
 Adding those satellites that are stripped by at least two thirds of their original stellar mass
 to the building blocks (to include also the main progenitors of spheroids C and D) did not significantly change this figure. 
 Right panel: The same mass-weighting is used to show the 30$-$50$-$70 percentiles of haloes A$-$E 
 virial masses of building blocks as a function of lookbacktime (black line with gray band) and those of surviving satellites
 (green line with lightgreen band). Again, the zoom-in panel shows the first Gyr of galaxy formation.
}
 \label{fig:6} 
\end{figure*}

In Figure~\ref{fig:4}, we show the average metallicities ($\log [\mathrm{Z}_\mathrm{stars}/\mathrm{Z}_\odot]$) 
of these satellites (blue circles), as a function of their stellar mass, and compare their values to those of the 
fully disrupted building blocks of the spheroids (yellow squares). 
The satellites / building blocks with less than or equal to 4 star formation 
snapshots (SFS) are plotted with open symbols in Figure~\ref{fig:4}. These indeed cover almost all galaxies
below metallicities~$< -3$. Because of the neglect of pre-enrichment from the very first generation of stars,
we do not trust the physical nature of the second line at metallicities~$\sim -4$ below the physical mass-metallicity relation 
which is starting at metallicities~$\sim -2$ (almost all symbols on this line are open).
All satellites shown here are those within a 280 kpc radius from the centre of the central galaxy in our simulations, 
which is a proxy for the virial radius of the Milky Way \citep{Koposov:2008}. 
Only a few satellites that are still bound to the central galaxy in our simulations
can be found outside of this radius.
Observed Milky Way satellites [Mg/H] values, corrected to better compare with our model calculations
(as described at the end of Section~\ref{sec:2}) within the same radius are plotted as red triangles with error bars, 
which for most galaxies fall within the symbol size. Values are taken from \citet{McConnachie:2012}.

Figure~4 does not evidence much difference in the metallicities of 
the building blocks and those of the surviving satellites at a given stellar mass. 
In order to see if the two populations could be drawn from the same underlying distribution, 
we conducted a Kolmogorov-Smirnov (KS) test, thereby excluding the open symbols and the galaxies in the shaded areas, 
where we suspect the resolution of our simulation to limit the robustness of our results.
Furthermore, we removed the mass dependence of the $\log [\mathrm{Z}_\mathrm{stars}/\mathrm{Z}_\odot]$ values 
by subtracting a quadratic polynomial fit from the data: 
$\log [\mathrm{Z}_\mathrm{stars}/\mathrm{Z}_\odot]= ax^2 + bx + c$ with $x = \log$ (Stellarmass $/M_\odot$), 
$a = -0.0208$, $b = 0.682$ and $c = -4.79$.
The resulting $D$ statistic of 0.09 and $p$-value of 0.38 indicate that 
the populations are consistent with each other. 
For this case, we conclude that the two populations follow the same underlying distribution.

Figure~\ref{fig:5} shows the average SFR (in $M_\odot$/yr) versus stellar mass, with the same
color coding as Figure~\ref{fig:4}.
We have calculated the average SFR as the sum of the SFRs prior to infall into
the halo of a larger galaxy (i.e. the galaxy was not yet a satellite) 
divided by the total number of timesteps during which there was star formation in this period.

It is clear from Figure~\ref{fig:5} that the average SFRs of the building blocks are typically higher at a given stellar 
mass than those of the surviving satellites. To quantify this difference, we again conducted a KS test on the two populations 
(open symbols, non-shaded area only) after subtracting a quadratic polynomial fit to remove the mass dependance
of the SFR values: $y = ax^2 + bx + c$ with $x = \log$ (Stellarmass $/M_\odot$), 
$y = \log$ (average SFR $/(M_\odot/\mathrm{yr})$) $a = 0.0479$, $b = -0.0509$ and $c = -3.83$.
We find a $D$ statistic of 0.49 and a $p$-value of $2.7 \cdot 10^{-22}$. 
To check the sensitivity of this strong result on our choices of calculating the average SFRs described earlier, 
we additionally calculate the averages in various different ways: 
the sum of the SFRs in all timesteps during which there is star formation 
divided by the number of timesteps in which there is star formation ($D$: 0.51, $p$-value: $4.0 \cdot 10^{-24}$); 
and the total stellar mass of the building block/satellite divided by the timespan of star formation 
($D$: 0.64, $p$-value: $5.7 \cdot 10^{-37}$). 
Furthermore, we compared the peak SFRs of the building blocks and the surviving satellites, which show the same trend 
again ($D$: 0.49, $p$-value: $8.0 \cdot 10^{-23}$).
Because the $p$-value is very low in all of these cases, we are convinced that our conclusion that 
the SFR in the building block population is higher than that in the surviving satellites is robust.

The left panel of Figure~\ref{fig:6} shows the SFR of all fully disrupted building blocks weighted by their stellar mass at that time of star formation,
versus the mass-weighted SFR of the surviving satellites, calculated separately for five of the Aquarius haloes, after which
the median is plotted with a black line for the building blocks and with a blue line for the satellites. 
With a coloured band, the 30$-$70 percentiles are shown.
The Aquarius simulation F is not included here, because it has a different time step size. 
The figure thus mainly represents the SFRs of the brightest objects
in these simulations, which are most easily detected. 
We see that the SFRs are especially different at early times (i.e. in the first Gyr, which is shown in the zoom-in panel). 
This may have important implications, for example the contribution to the reionization of the local universe.
The surviving satellites have not contributed significantly compared to the fully disrupted ones
and we might no longer be able to see host galaxies of the sources that reionized the local universe
\citep[see also][]{Weisz:2014,Boylan-Kolchin:2014,Boylan-Kolchin:2015}.
Furthermore, our result implies that globular clusters may be forming more easily in building block galaxies than in galaxies 
that survive as satellites until the present day, since they are thought to be outcomes of large starbursts occurring 
in the early universe \citep{Kruijssen:2012}.

\begin{figure*}
 \includegraphics[width=\textwidth]{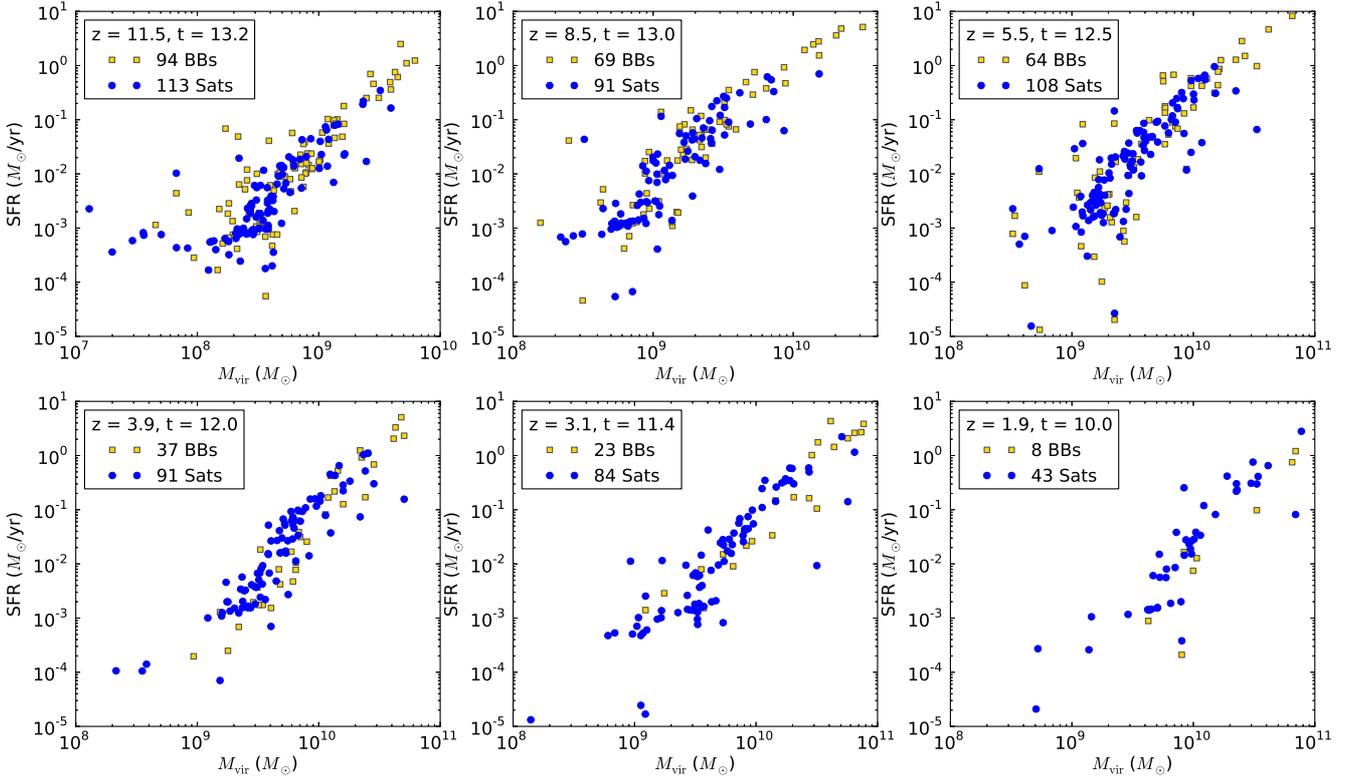}
 \caption{Virial mass versus SFR for the galaxies that become building blocks (yellow boxes) and the ones
 that survive as satellites (blue circles), at six different time steps. 
 The time and redshift labels are shown in the upper left corner of the panels.
 Also the total number of objects at that time step is indicated in the legend of each panel.}
 \label{fig:12}
\end{figure*}

To shed some light on the origin of the difference in SFR between building blocks and surviving satellites, we show
in Figure~\ref{fig:12} the virial mass versus SFR for all galaxies that still have their own dark matter halo 
at six different time steps in the early universe. Galaxies that end up as building blocks are marked as yellow squares,
whereas galaxies that survive as a satellite galaxy to the present day are visualized as blue circles. 
From this figure, it is clear that until $\sim$11.5 Gyr ago, the most massive galaxies are those that are building 
blocks at the present epoch. These building blocks are thought to be associated with high-density peaks that collapsed at higher redshifts, 
compared to the present-day surviving satellites that descended from more average density fluctuations in the early universe
\citep{Barkana:2001,Diemand:2005}. At early epochs these satellites have lower virial masses than the building blocks, 
resulting in longer timescales for merging with the central galaxy and lower SFRs.
This same trend is shown in the right panel of Figure~\ref{fig:6} where we apply the same mass weighting as we did to show
the median SFR of the brightest objects at a particular lookbacktime, this time to show the median virial mass of the same
objects (as well as the 30$-$70 percentiles again). The fact that the curves in the right panel are decreasing after some time, 
is due to tidal disruption of the haloes. Note that at the same time ($\sim$11 Gyr ago) the SFR of the building blocks 
drops below that of the satellites (left panel of Figure~\ref{fig:6}) as their virial mass drop below that of the satellites
(right panel of Figure~\ref{fig:6}). 

\begin{figure*}
 \includegraphics[width=\textwidth]{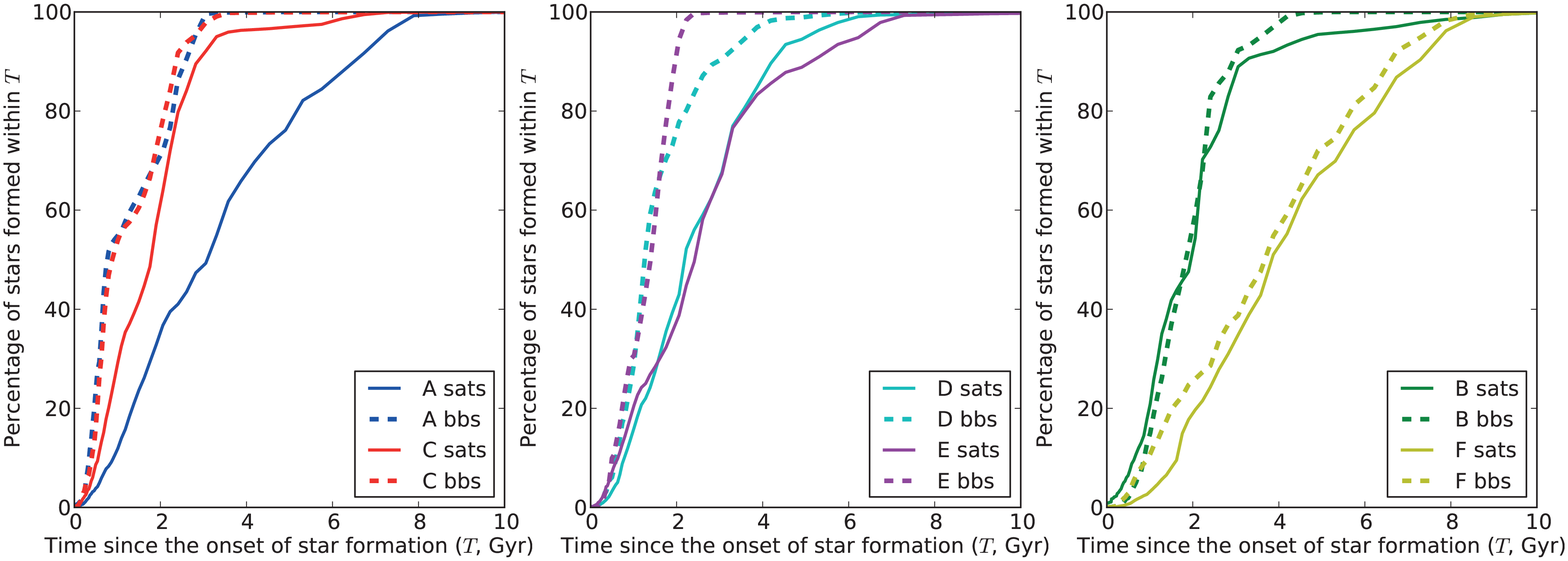}
 \caption{Percentage of stars in surviving satellites (solid lines) versus building blocks (dashed lines)
 that is formed since the onset of star formation. For spheroids A and C (left panel) the difference between the
 two populations is very large already after $\sim 1$ Gyr, for spheroids D and E (middle panel) 
 the difference is largest after $\sim 2$ Gyr, whereas for spheroids B and F (right panel) the difference is negligible.
 All building blocks that form the stellar spheroid are included, also the stripped material from surviving satellites.}
 \label{fig:7}
 \vspace{0.4cm}
 \includegraphics[width=\textwidth]{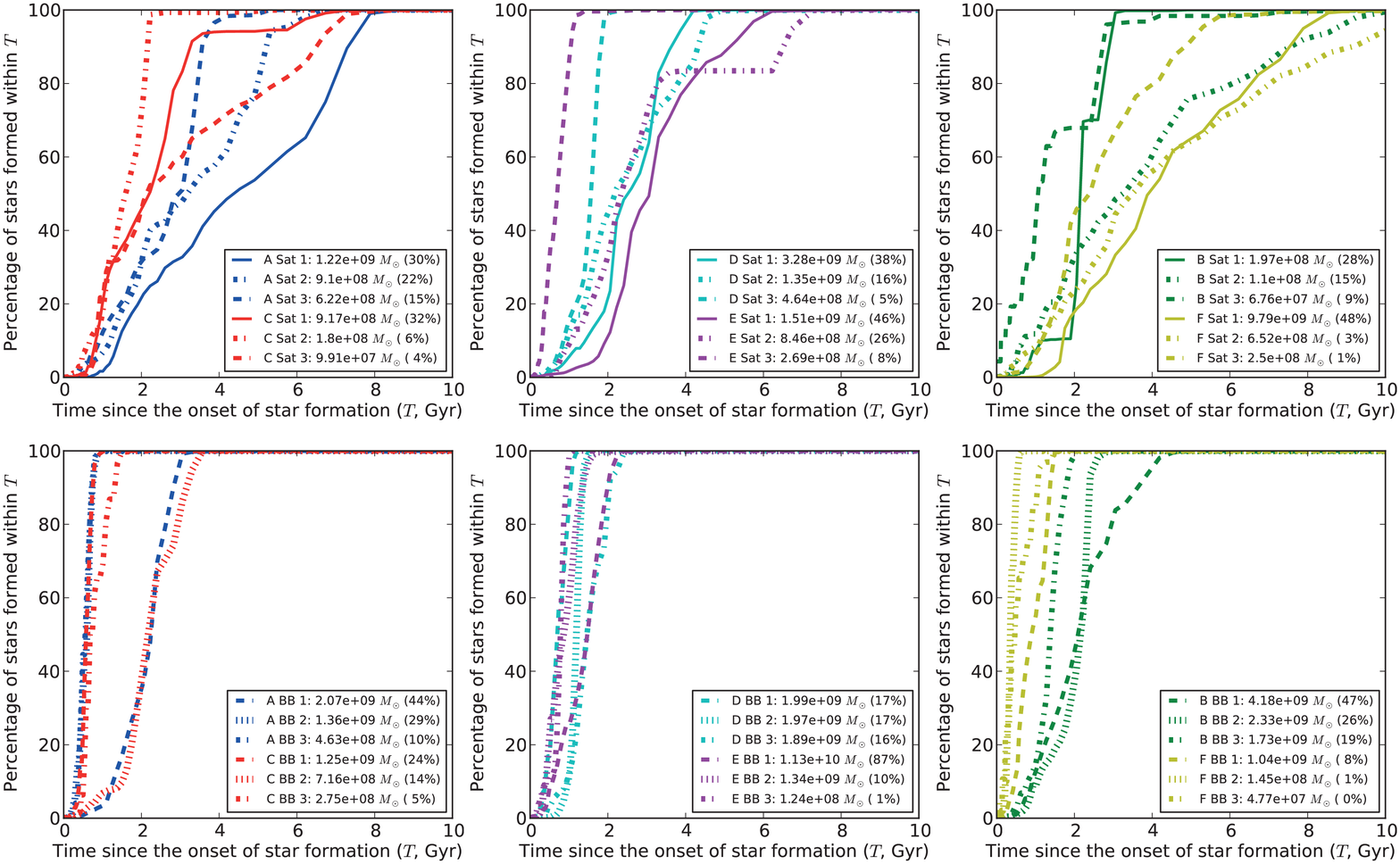}
 \caption{Percentage of stars formed since the onset of star formation in the three most massive surviving satellites 
 that have lost less than 20\% of their stars (top three panels) and the three most massive building blocks 
 that have been fully disrupted (bottom three panels). As in Figure~\ref{fig:7}, spheroids A and C are visualized 
 in the left panels, spheroids D and E in the middle panels, and spheroids B and F in the right panels. Each color indicates a seperate
 spheroid, the same colorcoding is used as in Figure~\ref{fig:7}. The percentage of the total mass in satellites/building blocks 
 that is contained in a particular satellite/building block is given in brackets in the legend of each panel.
 The most massive surviving satellites shown here (numbers 1) of spheroids B (indicated with the green solid line in the top right panel),
 D (the cyan solid line in the top middle panel) and E (magenta solid line in the top middle panel) lost respectively 6\%, 2\%
 and 2\% of their initial mass through tidal stripping, the other satellites did not lose any mass. }
 \label{fig:8}
\end{figure*}

\section{Comparison of Building Blocks and Surviving Satellites through the Star Formation Timespan}\label{sec:5}

\subsection{Timescales of growth}\label{sec:5a}

\begin{figure}
 \includegraphics[width=\columnwidth]{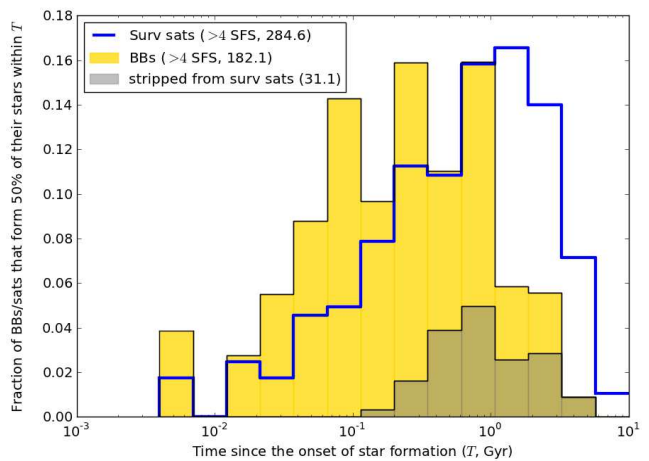}
 \caption{Fraction of building blocks (filled gold histogram) and surviving satellites (transparent histogram with thick blue edge) with more than 4
 SFS that form 50\% of their stars within the time since the onset of star formation ($T$, Gyr), for the six Aquarius haloes combined. 
 ``Building blocks'' that are stars stripped from surviving satellites are divided among the two populations according to the mass fraction 
 of the initial satellite mass. The dark shaded filled histogram shows how this material contributes to the building block distribution. 
 The total number of building blocks and surviving satellites in these distributions are shown in brackets in the legend. }
 \label{fig:9}
\end{figure}

\begin{figure*}
\includegraphics[width=\textwidth]{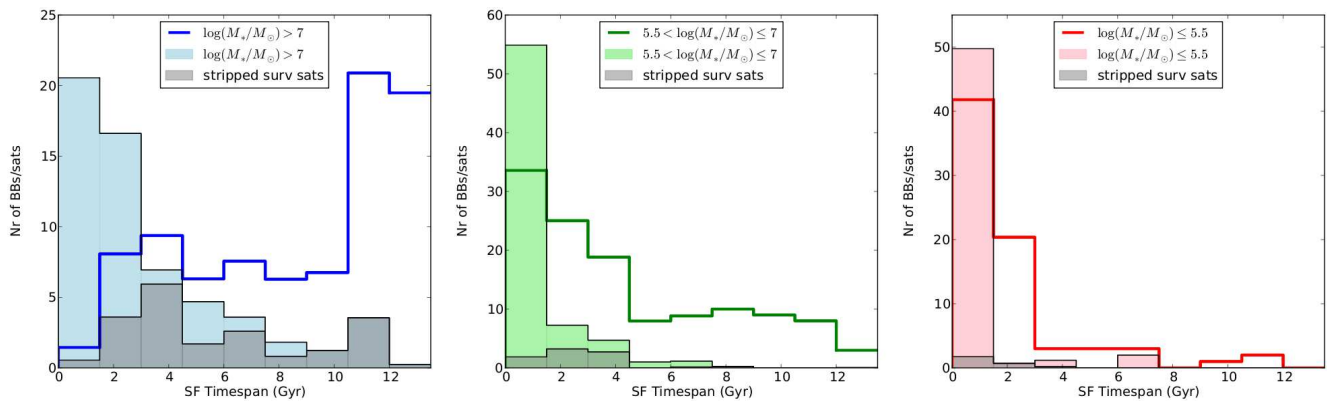}
 \caption{Number of building blocks (filled histograms) and surviving satellites (transparent histograms with thick edges)
 with more than 4 SFS per timespan it took to form all their stars (SF Timespan, binned in bins of 1.5 Gyr). The populations
 are split up into three different stellar mass regimes. High stellar mass building blocks/satellites ($M_{*} > 10^7 M_\odot$)
 are shown with blue colors in the left panel, intermediate stellar mass building blocks/satellites ($10^{5.5} < M_{*}/M_\odot \leq 10^7$) with 
 green colors in the middle panel, and low stellar mass building blocks/satellites ($M_{*} \leq 10^{5.5} M_\odot$) with red colors on the right.
 With a dark shading, the material stripped from surviving satellites is indicated again.
 }
 \label{fig:10}
\end{figure*}

As already shown in the previous section, we find in our simulations that the average SFR in the surviving satellites 
is lower than in building blocks of the same mass (see Figure~\ref{fig:5}), so it generally takes them 
a longer time to form their stars. This is visualised in more detail in Figures~\ref{fig:7} to \ref{fig:10}.

The mass build-up of the accreted stellar spheroids (in percentage of the total accreted mass over time) 
since the beginning of star formation is shown for both building blocks and surviving satellites in Figures
\ref{fig:7} and \ref{fig:8}. For each galaxy, the moment at which the first stars in that galaxy begin to form 
is set as the zero Time $T$. 
If a surviving satellite or building block experienced a merger with another galaxy, the star formation histories of the two
galaxies are added together. The galaxy after the merger is therefore always considered to have started forming stars as early as the
earliest of the two galaxies started to form stars. We do not examine the complete merger histories of building blocks and
satellites here, but \citet{Deason:2014a} estimated that $\sim$10\% of the dwarf galaxies with a stellar mass $>10^6 M_\odot$ 
that are within the virial radius of the host experienced a major merger since $z=1$.

Compared with the lookbacktime (used the previous sections in Figures \ref{fig:1}, \ref{fig:2}, \ref{fig:3} and
\ref{fig:6}) the onsets of star formation are shifted so that they all start at the same zero time, 
thus allowing us to address the enrichment histories of the building blocks and satellites on a similar internal time scale. 
This is particularly relevant since several enrichment processes have longer delay time than others; e.g., 
SNe type Ia originate from lower mass progenitors than SNe type II and thus the former can only 
contribute significantly to enrich the galaxy after a certain time since the onset of star formation. 
The relative contributions of these SN types will have their imprint on the chemical abundances of the next generations of stars; 
whereas SNe type II are producing lots of $\alpha$-elements such as Mg, Ca and Ti, SNe Ia are thought to be the main contributors for Fe for instance.
The shifts that we applied, from the universal time to $T$, are no more than 0.1 Gyr for the most massive building blocks 
(of which we plotted the SFR in Figure~\ref{fig:2}) and no more than 0.5 Gyr for the three most massive satellites of each spheroid.

We see that in spheroids A and C (leftmost panel of Figure~\ref{fig:7}) $\sim 50\%$ of the stars in building blocks
are formed within one Gyr, whereas only $10 - 30 \%$ of the stars in the surviving satellites are formed during
this timespan. For spheroids D and E (middel panel), the difference between the two percentages is the largest after 
approximately two Gyr, when $\sim 70 - 80\%$ of the stars in the building blocks were formed, compared to $\sim 40\%$ 
in the satellites. For spheroids B and F (right panel) however, the difference between the build-up of the two populations
of stars is much smaller.

In Figure~\ref{fig:8} we show the same growth rates for the surviving satellites (top panels) and the building blocks
(bottom panels) as in Figure~\ref{fig:7}, but now split out into contributions from three massive progenitors/surviving
satellites in the population. As in Figure~\ref{fig:7} spheroids A\&C, D\&E, and B\&F are shown from left to right.
The solid lines in the top panels and the dashed lines in the bottom panels correspond to the most massive satellites 
that have lost less than 20\% of their stars and most massive building blocks that have been fully disrupted respectively, 
the second and third most massive galaxies are shown with other linestyles (see legend). 
The contribution of that particular object to the total stellar mass in satellites or building blocks
is given in between brackets in the legend of each panel. This gives an estimate of the weighting of that line with respect to 
the total build-up of satellites or building blocks over time (Figure~\ref{fig:7}).

From Figure~\ref{fig:8} we learn that the spread in stellar mass build-up over time for massive satellites is larger than 
for massive building blocks. Comparing Figures \ref{fig:7} and \ref{fig:8}, we see that the most massive surviving 
satellites (shown with solid lines in the top three panels of Figure~\ref{fig:8}) 
resemble quite well the total build-up of the satellites in that spheroid over time (solid lines in Figure~\ref{fig:7}).
The same is true for the most massive building blocks of spheroids B and E, i.e. the dashed lines in the bottom panels 
of Figure~\ref{fig:8} match the dashed lines in Figure~\ref{fig:7} for these spheroids quite well. 
For spheroid A on the other hand, the second most massive building block shown in Figure~\ref{fig:8} forms
many more stars in a short period of time than the most massive one, thereby increasing the total percentage of stars 
formed early on. For the other spheroids (in particular spheroid F), there is also a discrepancy between the 
dashed lines in the bottom panels of Figure~\ref{fig:8} and those in Figure~\ref{fig:7} because the
most massive progenitor is not fully disrupted and therefore not included in Figure~\ref{fig:8}.

Focussing now on the time span during which the first fifty percent of the stars in a building block/surviving satellite
are formed, we show an overview of this distribution in Figure~\ref{fig:9}. From this it is once again clear that the building blocks 
form the first 50\% of their stars in a shorter time than the surviving satellites on average, as we expected from Figures~\ref{fig:7} and \ref{fig:8}. 
In Figure~\ref{fig:9} the time since the onset of star formation ($T$, Gyr) is binned in 15 equal-size bins on a logarithmic scale from $10^{-2.4}$ to 10 Gyr. 
We left out the building blocks/surviving satellites that had star formation in less than or equal to 4 snapshots again, 
which would otherwise pile-up in the lowest bin.
The contribution from stripped surviving satellites is divided among the two populations according 
to the mass fraction. For example, the most massive progenitor of spheroid D is counted as 0.94 building block 
and 0.06 surviving satellite. With a dark shading in the building block histogram, we show that this material follows a 
different distribution than the total building block population and is more similar to that of the surviving satellites.

Finally, in Figure~\ref{fig:10} we split the total number of building blocks/surviving satellites up into three mass regimes, i.e.
massive ($> 10^7 M_\odot$, in blue), intermediate mass ($10^{5.5} < M/M_\odot \leq 10^7$, in green) and low-mass ($\leq 10^{5.5} M_\odot$, in red) 
shown from left to right respectively. For the building blocks, indicated again with filled histograms, the fraction of stars 
in the lowest time bin relative to the total in that mass regime becomes larger if they are less massive, 
thus the chance that they formed their stars in a short timespan becomes larger with decreasing mass.
However, the shape of the distribution is peaking at the shortest timespan in all three mass regimes for the building blocks. 
The opposite is true for the surviving satellites: the probability density function of the massive satellites peaks at large timespans, 
and the peak moves towards shorter timespans if the satellites become less massive, towards a distribution that is similar in shape 
to that of the building blocks for the lowest mass satellites. The ``building blocks'' that were stripped
from surviving satellites are indicated again with a dark shading, and as in Figure~\ref{fig:9}, they are 
divided among the two populations according to the mass fraction of the initial satellite mass.

The three stellar mass bins chosen in Figure~\ref{fig:10} represent the various classes of dwarf satellite galaxies surrounding the Milky Way 
(as illustrated in comparison with Figure \ref{fig:4}). From the observed trends with stellar mass we can conclude that the population 
of the faintest (also called ``ultra-faint'') dwarfs show slightly more overlap in their star formation properties with the building blocks of similar mass 
than the brighter ``classical'' dwarf galaxies.  

\subsection{Relation between Iron Abundance and Supernovae Type Ia Delay Time Distribution}\label{sec:5b}

One key observable of our Milky Way system is that although its halo population is $\alpha$-rich \citep{Hawkins:2014,Jackson-Jones:2014}, 
its (classical) satellites are predominantly $\alpha$-poor, with an exception for their most metal-poor components 
\citep[e.g.,][]{Shetrone:1998,Shetrone:2001,Shetrone:2003,Tolstoy:2003,Venn:2004,Koch:2008,Kirby:2010, Starkenburg:2013a,Jablonka:2015, Frebel:2015}. 
At metallicities of [Fe/H]$\sim -1$, stars in the local halo have [$\alpha$/Fe]  ratios ([$\alpha$/Fe]$\sim 0.2-0.4$) 
that are approximately $0.2-0.6$~dex higher than those in dwarf galaxies in the Local Group \citep[see for a review][]{Tolstoy:2009}.

Several modelling efforts have already pointed out that indeed such a discrepancy could arise in a stellar halo built out of few early-accreted, 
massive main progenitor galaxies. These would have had high SFRs and have been enriched primarily by type II SNe, whereas the surviving satellites
had lower SFRs over a longer period of time, during which also type Ia SNe contributed to their metal content, resulting in a lower [$\alpha$/Fe] 
abundance \citep{Robertson:2005,Font:2006,Font:2006a,Geisler:2007}.  If type Ia SNe start to contribute significantly only after a certain timescale,
this causes the [Fe/H] ratio to increase with respect to [$\alpha$/H], leading to a ``knee'' in the [Fe/H] versus [$\alpha$/Fe] diagram. Support for 
such a scenario comes from the position of the observed knee in various galaxies; the $\alpha$-element knee of the Sculptor dSph for example is estimated 
to be around [Fe/H]$\approx -1.8$ \citep{Tolstoy:2009}, whereas for the more massive Sagittarius it takes place at [Fe/H]$\approx -1.3$ \citep{de-Boer:2014}. 
This is consistent with our Figure \ref{fig:10}, where we see that the higher mass satellites have a higher SF efficiency, especially at earlier times. 

As discussed before, our semi-analytic model does not include finite stellar lifetimes, but is based on an instantaneous recycling approximation, 
and therefore the relative contribution from type Ia SNe and type II SNe are not modelled directly. However, from the distribution of star formation timespans in
the various building blocks and satellites as shown in Figures \ref{fig:7} until \ref{fig:10} we can still infer some information on the [$\alpha$/Fe] ratios expected. 
For instance, by making the (very rough, but common) assumption that all spheroid stars formed after a 
certain time $T$, say 1~Gyr, are significantly enriched in iron from type Ia SNe, and those before that time are $\alpha$-rich, 
we can use Figure~\ref{fig:7} to predict that of the six Aquarius haloes, spheroids A and C have the most dominant
high-$\alpha$ populations, spheroids B and F the least. 
\citet{De-Lucia:2014,De-Lucia:2015} have investigated the implementation of different delay time distributions for SN Ia explosions  
within a semi-analytical model that includes chemical evolution, but which is otherwise very similar to ours. 
They conclude that the Milky Way stellar disc metallicity distribution function is best represented for delay time distributions that are fairly
broad, rather than strongly peaked at either short or intermediate delay times. In all these cases, the effective [O/Fe] yield 
(for a simple stellar population at fixed metallicity) is at a level $\sim$0.25 dex lower and most steeply dropping around a 1 Gyr timescale, 
consistent with our simple assumption.

We see from Figure \ref{fig:9} that 99\% of the building blocks and 92\% of the surviving satellites form the first 50\% of their stars in 3.3 Gyrs.
The observed discrepancy in [$\alpha$/Fe] values between the stellar halo and the satellites of the Milky Way would therefore not be expected to show up 
in any of our modelled haloes, making 3.3 Gyr an upper limit on the delay time scale for SNe Ia to become significantly abundant in the chemical enrichment process.
Should on the other hand the relevant contribution delay time be as small as $6.5\cdot 10^{-2}$~Gyr, then only 19\% of the building blocks would be $\alpha$-rich, 
versus 10\% of the satellites. A difference that small would also be inconsistent with observations, making $6.5\cdot 10^{-2}$~Gyr a lower limit on this time scale.

We would like to note that although the exact distribution of type Ia's as a function of time is highly uncertain \citep[see e.g.,][]{Matteucci:2009, Maoz:2014}, 
many authors find that the delay time distribution has a power-law form, $\propto t^{-1}$, according to which $\sim$50\% of the type Ia SNe occur within $\sim$1~Gyr. 
The time scale $T$ discussed here represents a cumulative result of all SNe Ia explosions on the galaxy enrichment, capable of moving from a regime of forming 
predominantly high-$\alpha$ stars to low-$\alpha$ stars, not the time scale at which these stars start exploding. 
In the solar neighbourhood the change of the [O/Fe] slope around [Fe/H]$= -1$ in the [O/Fe] versus [Fe/H] diagram is consistent with galactic chemical evolution models 
in which the overall delay time scale for significant SNe Ia enrichment is $\sim$1 Gyr \citep[eg.][]{Matteucci:2001}. These timescales could be different in regions 
with a different SFR, such as the bulge or the outer disc \citep[eg.][]{Pipino:2009}.

The [$\alpha$/Fe] distribution function of surviving satellites versus that of the inner halo was already modelled in detail by \citet{Font:2006}, 
who also concluded that the bulk of the halo formed from massive satellites accreted early on. As noted by \citet{Font:2006a} and \citet{Johnston:2008}
we see that in our models the net efficiency of star formation - and in particular the difference therein between the building blocks and satellites - 
shows some variations between the modelled Milky Way-like systems. This means that, largely independent of the assumptions made for the exact delay time distributions and/or 
Fe-enrichment mechanisms, we might expect to see different [$\alpha$/Fe] distributions in various Milky Way-mass systems depending on their detailed formation histories. 
From an observational point of view, this is an exciting prospect offering us a different angle by which the history of a stellar halo can be unravelled. 
Currently, we do not have a clear picture on the [$\alpha$/Fe] ratio in external Milky Way-like haloes (although \citet{Vargas:2014} measured [$\alpha$/Fe] 
abundances of four stars in the outer halo of the Andromeda Galaxy and found them $\alpha$-enriched) but this will very likely change in the era of the E-ELTs \citep[see e.g.,][]{Battaglia:2011}.

\section{Conclusion}\label{sec:6}

In this paper we have investigated the accreted stellar spheroids of Milky Way like galaxies with the
Munich-Groningen semi-analytical model of galaxy formation, combined with the high-resolution Aquarius
dark matter simulations. Typically, each of the accreted spheroids was built by only a few main progenitor galaxies
and the majority of stars that end up in our Milky Way like stellar spheroids is 10$-$13 Gyr old. 
In three of our six galaxies (C, D and F) a large fraction of the spheroid stars is stripped from satellites that are 
surviving to the present day. For spheroids C\&D these may be resembling the Sagittarius dwarf's contribution to the Milky Way halo.
Spheroid F is atypical as a Milky Way analog because it accreted $\sim 10^{10} M_\odot$ in stars over the last $\sim 3$~Gyr.

We compared the properties of the building blocks of the Milky Ways stellar spheroid
to those of the surviving satellites and found that in terms of the stellar mass $-$ metallicity relation, 
the difference between the two populations is small, but that the former have significantly higher star formation rates on average - 
they form comparable amounts of stars in a shorter time (see Figures \ref{fig:6} and \ref{fig:10}). 
In particular, the more massive surviving satellites show a larger variety in stellar mass build-up over time than the massive building blocks (Figure~\ref{fig:8}). 
On the other hand, the faintest surviving satellites build up mass in a similar fashion to building blocks with similar mass (right panel of Figure~\ref{fig:10}). 

From these results, we expect the stellar spheroid to be more enriched in $\alpha$-elements compared to Fe than the surviving satellites, 
as we observe in the Milky Way system. However, a quantitative analysis of the detailed chemical evolution will require a more sophisticated model 
and accurate descriptions of the delay time for SNe type Ia. Furthermore, we are dealing with a stochastic process since we are comparing the 
spheroids of only six Milky Way-like galaxies, that have accreted components which are dominated by a few objects.
This results in some of the Aquarius haloes having a better match with the Milky Way galaxy in terms of overall stellar mass 
and spheroid metallicity, while others have an accretion history that more closely matches that of the Milky Way.
Also, we observe some scatter from system to system in our models of the timescale of star formation in satellite galaxies 
and the timescale of star formation in the main halo. A prediction of these models is therefore that not all Milky Way-mass systems 
will show [$\alpha$/Fe] ratios similar to those in the Milky Way.

\section*{Acknowledgements}
The authors are indebted to the Virgo Consortium, which was responsible for designing and running the halo simulations of the Aquarius Project 
and the L-Galaxies team for the development and maintenance of the semi-analytical code. In particular, we are grateful to Gabriella De Lucia and 
Yang-Shyang Li for the numerous contributions in the development of the code.
Furthermore, we would like to thank the anonymous referee for valuable comments that helped to improve this paper.
PvO thanks the Netherlands Research School for Astronomy (NOVA) for financial support.
ES gratefully acknowledges funding by the Emmy Noether program from the Deutsche Forschungsgemeinschaft (DFG).
AH acknowledges financial support from a VICI grant.

\bibliographystyle{mnras}
\bibliography{paper} 

\label{lastpage}
\end{document}